\documentclass[12pt,preprint]{aastex}

\slugcomment{}

\shorttitle{The Importance of Non-Equilibrium Chemistry}
\shortauthors{Leggett et al.}

\begin{document}


\title{Near-Infrared Spectroscopy of the Y0 WISEP J173835.52$+$273258.9 and the Y1 WISE J035000.32$-$565830.2: \\ the Importance of Non-Equilibrium Chemistry}


\author{S. K. Leggett\altaffilmark{1}}
\email{sleggett@gemini.edu}
\author{P. Tremblin\altaffilmark{2,3}}
\author{D. Saumon\altaffilmark{4}}
\author{M. S. Marley\altaffilmark{5}}
\author{Caroline V. Morley\altaffilmark{6}}
\author{D. S. Amundsen\altaffilmark{3}}
\author{I. Baraffe\altaffilmark{3,7}}
\author{G. Chabrier\altaffilmark{3,7}}

\altaffiltext{1}{Gemini Observatory, Northern Operations Center, 670
  N. A'ohoku Place, Hilo, HI 96720, USA} 
\altaffiltext{2}{Maison de la Simulation, CEA-CNRS-INRIA-UPS-UVSQ, USR 3441, Centre d'\'etude de Saclay, F-91191 Gif-Sur-Yvette, France}
\altaffiltext{3}{Astrophysics Group, University of Exeter, EX4 4QL Exeter, UK}
\altaffiltext{4}{Los Alamos National Laboratory, PO Box 1663, MS F663, Los Alamos, NM 87545, USA}
\altaffiltext{5}{NASA Ames Research Center, Mail Stop 245-3, Moffett Field, CA 94035, USA}
\altaffiltext{6}{Department of Astronomy and Astrophysics, University of California, Santa Cruz, CA 95064, USA}
\altaffiltext{7}{Ecole Normale Sup\'erieure de Lyon, CRAL, UMR CNRS 5574, F-69364 Lyon Cedex 07, France}

\begin{abstract}
We present new near-infrared spectra, obtained at Gemini Observatory, for two Y dwarfs:  WISE J035000.32$-$565830.2 (W0350) and WISEP J173835.52$+$273258.9 (W1738). A FLAMINGOS-2   $R = 540$ spectrum was obtained for W0350, covering $1.0 < \lambda ~\mu$m $< 1.7$, and a cross-dispersed GNIRS   $R = 2800$ spectrum was obtained for W1738, covering 
0.993 -- 1.087  ~$\mu$m,  1.191 -- 1.305  $\mu$m, 1.589 -- 1.631    ~$\mu$m, and
1.985 -- 2.175 ~$\mu$m,  in four orders. We also present revised $YJH$ photometry for W1738, using new NIRI $Y$ and $J$ imaging, and a re-analysis of the previously published NIRI $H$ band images. We compare these data, together with previously published data for late-T and Y dwarfs, to cloud-free models of solar metallicity, calculated both in
chemical equilibrium  and with disequilibrium driven by vertical transport. We find that for the Y dwarfs the non-equilibrium models reproduce the near-infrared data better than the equilibrium models. The remaining discrepancies suggest that fine-tuning the CH$_4$/CO and NH$_3$/N$_2$ balance is needed. Improved trigonometric parallaxes would improve the analysis. 
Despite the uncertainties and discrepancies, the models reproduce the observed near-infrared spectra well.  
We find that for the Y0, W1738,   $T_{\rm eff} = 425 \pm 25$K and log $g = 4.0 \pm 0.25$, and for the Y1, W0350,  $T_{\rm eff} = 350 \pm 25$K and log $g = 4.0 \pm 0.25$. W1738 may be metal-rich. Based on evolutionary models, these temperatures and gravities correspond to a mass range for both Y dwarfs of 3 -- 9 Jupiter masses, with W0350 being a cooler, slightly older, version of W1738;
the age of W0350 is 0.3 -- 3  Gyr, and the age of W1738 is 0.15 -- 1 Gyr.

\end{abstract}


\section{Introduction}

The atmospheres of giant gaseous planets and brown dwarfs (objects with a mass below that required for stable nuclear fusion, mass $\lesssim 80$ Jupiter masses) are molecule-rich and chemically complex. The deep atmosphere is fully convective; 
there can be detached convection zones above the radiative-convective boundary, if the pressure and composition are such that there is strong absorption at the wavelength typical of the flux being emitted at the temperature of that particular layer.  The reader is referred to the review  by Marley \& Robinson (2015) for further discussion.
The brown dwarf atmospheres are turbulent, and chemical species are mixed vertically through the atmosphere. Mixing occurs in the convective zones, but may also occur in the  nominally quiescent radiative zone by processes such as gravity waves (e.g. Freytag et al. 2010). If mixing occurs faster than local chemical reactions can return the species to local equilibrium, then abundances can be very different from those expected for a gas in equilibrium. Species whose abundances are significantly impacted by mixing in brown dwarf atmospheres are CH$_4$, CO, CO$_2$, N$_2$ and NH$_3$ (e.g. Noll, Geballe \& Marley 1997, Saumon et al. 2000, Golimowski et al. 2004, Leggett et al. 2007, Visscher \& Moses 2011, Zahnle \& Marley 2014). 

Also, various species condense, forming cloud decks in the photosphere. For L dwarfs with effective temperature 
$1300 \lesssim T_{\rm eff}$~K $\lesssim 2300$ (e.g. Stephens et al. 2009) the condensates are composed of iron and silicates, and for T dwarfs with 
$500 \lesssim T_{\rm eff}$~K $\lesssim 1300$ (e.g. Morley et al. 2014)
they consist of  chlorides and sulphides 
(e.g. Tsuji et al. 1996, Ackerman \& Marley 2001, Helling et al. 2001, Burrows et al. 2003, Knapp et al. 2004,
Saumon \& Marley 2008, Stephens et al. 2009, Marley et al. 2012, Morley et al. 2012, Radigan et al. 2012, Faherty et al. 2014). As 
$T_{\rm eff}$ decreases further,
the next species to condense are calculated to be H$_2$O for $T_{\rm eff} \approx$ 350~K and 
NH$_3$ for  $T_{\rm eff} \approx$ 200~K 
(Burrows et al. 2003, Morley et al. 2014). 

It had been hoped that the chemistry of the atmospheres of very cold brown dwarfs, those with  $400 \lesssim T_{\rm eff}$~K $\lesssim 800$,  would be simpler than it is for the warmer L and T dwarfs, as many species will have condensed out, and water clouds will not yet have formed. Twenty-three brown dwarfs are now known with effective temperature $T_{\rm eff} \le 500$~K, and these have been classified as Y dwarfs   (e.g. Kirkpatrick et al. 2012, Leggett et al. 2013). It turns out that modeling their atmospheres is not simple (e.g. Leggett et al. 2015, hereafter L15).

All but one of the known Y dwarfs have been found using the {\it Wide-field Infrared Survey Explorer} ({\it WISE}; Wright et al. 2010) by\,: Cushing et al. (2011, 2014); Kirkpatrick et al. (2012); Liu et al. (2012); Luhman (2014); Pinfield et al. (2014); 
Schneider et al. (2015); Tinney et al. (2012). The remaining object was discovered by 
Luhman, Burgasser \& Bochanski (2011) as a companion to a white dwarf, using images from the Infrared Array Camera onboard the {\it Spitzer Space Telescope} (Werner et al. 2004).
L15 studied the properties of seventeen Y dwarfs using near-infrared photometry and spectroscopy, together with {\it WISE} and {\it Spitzer} mid-infrared photometry.  The observations were compared to spectra and colors generated from model atmospheres with a variety of cloud cover --- cloud-free models from Saumon et al. (2012, hereafter S12), models with homogeneous layers of  chloride and sulphide  clouds from Morley et al. (2012), and patchy cloud models from Morley et al. (2014). The models include updated opacities for NH$_3$ and pressure-induced H$_2$.  It was found that the models qualitatively reproduced the trends seen in the observed colors, and that the cloud layers are thin to non-existent for these brown dwarfs with  $T_{\rm eff} \approx  400$~K. However the model fluxes  were a factor of two low at the $Y$, $H$, $K$, [3.6], and W3(12~$\mu$m) bands. 
The models used in L15 all assumed equilibrium chemistry, and it was suggested that
much of the discrepancy could be resolved by significantly reducing the NH$_3$ abundance, perhaps by vertical mixing.

In this work we compare observed Y dwarf photometry and spectroscopy to models by Tremblin et al. (2015, hereafter T15) which include non-equilibrium chemistry, as well as an updated line list for CH$_4$ absorption. We also present new near-infrared spectroscopy for two Y dwarfs, and revised near-infrared photometry for one of these.

\section{Observations}

\subsection{WISE J035000.32$-$565830.2}

The discovery of WISE J035000.32$-$565830.2 (W0350) was published by Kirkpatrick et al. (2012), who classified it as a Y1 dwarf. Table 1 gives photometry and astrometry for this target, with source references.

We observed W0350 at Gemini South, via program GS-2014B-Q-17, using the  FLAMINGOS-2 imager and spectrometer (Eikenberry et al 2004). We obtained a $1.0 < \lambda ~\mu$m $< 1.7$ spectrum using the JH grism with JH blocking filter, and the    0$\farcs$72 slit. The resulting resolving power is  $R = 540$. 

On 2014 November 7, fifty two 300$\,$s frames were obtained over an approximately five hour period in photometric conditions and   0$\farcs$7 seeing. An ``ABBA'' offset pattern was used with offsets of $\pm10\arcsec$ along the slit. On 2014 December 4 
and 2015 January 3 sixteen and fourteen frames, respectively, were obtained in similar conditions with the same instrument configuration. The data from 2015 January 3 were taken at a significantly higher airmass (1.5 -- 1.9, compared to  1.2 -- 1.3 on the earlier two nights), and were not combined with the other datasets due to the lower signal to noise ratio (S/N). The 68 frames from November and December were reduced in a standard way using calibration lamps on the telescope for  flat fielding and wavelength calibration. The 68 300$\,$s images were combined using the {\tt gemcombine} IRAF routine, giving 5.7 hours on this source. 

Bright F3 and F7 dwarf stars were observed on each night, to remove telluric absorption features and flux calibrate the spectra. The stars used on 2014 November 7 were HD 13517 and HD 30526, and on  2014 December 4 HD 36636. Template spectra for these spectral types were obtained from the spectral library of Rayner et al. (2009), and used to determine a one-dimensional sensitivity function. A one-dimensional spectrum for the target was extracted from the combined image using the trace of the standard stars for reference. The sensitivity spectrum was then used to correct the shape of the target spectrum, and a final flux calibration was done on the target spectrum using the observed $J$ band photometry for the source ($Y$ band coverage was incomplete, and $H$ band was noisy). Figure 1 shows our spectrum smoothed with a 3 pixel boxcar. The uncertainty in the spectrum was determined by the sky noise, and is shown in Figure 1; S/N across the $J$ band peak is $\sim 10$, while across the $H$ band it is  $\sim 4$.

Schneider et al. (2015) used the {\it Hubble Space Telescope (HST)} Wide Field Camera 3 to obtain near-infrared slitless grism spectroscopy of W0350. The G141 grism was used, covering 1.10 -- 1.70 ~$\mu$m at R$\sim130$. Figure 1 shows both our Gemini spectrum and the {\it HST} spectrum, 
where the latter has been multiplied by a factor of 1.07 to bring it into agreement with our $J$ band photometry. The agreement is good across the  $J$ band but not as good across the $H$ band, where the flux peaks differ by $\sim 30$\%. Synthesizing the  $H$ band photometry and comparing it to the measured value suggests that our spectrum is too bright, while the Schneider spectrum is too faint. The discrepancy is likely
due to the faintness of the source and the resulting low S/N, although variability cannot be excluded. The compilation of photometric variability of brown dwarfs by Crossfield (2014) shows that T dwarfs can be variable at the $\sim 10$\% level.
We compare our spectrum to models in \S 3.3.

\subsection{WISEP J173835.52$+$273258.9}

The discovery of WISEP J173835.52$+$273258.9 (W1738) was published by Cushing et al. (2011), who classified it as a Y0 dwarf. Table 1 gives photometry and astrometry for this target, with source references.

The $YJH$ photometry for W1738 presented in Table 1 differs from that published by L15. As part of the Gemini North program GN-2013A-Q-21, W1738 was observed for several hours with the Gemini near infrared imager  (NIRI; Hodapp et al. 2003) in $Y$ and $J$ in a search for variability. The variability result will be published elsewhere; here we present improved values of $Y$ and $J$. The $J$ band result is significantly different from that previously published. We re-examined the $H$ band photometry obtained on the same night as the previously published data, which was a night of poor seeing. Re-reducing the earlier $H$ band data set paying closer attention to instances of very poor seeing, and therefore detections of low significance,
produces the revised $H$ value given in Table 1. The new $J$ and $H$ values are now in better agreement (within 2 $\sigma$) with the synthetic values determined by Schneider et al. (2015) from slitless $HST$ spectra.

We obtained spectroscopy for W1738 on Gemini North,  via program GN-2014A-Q-64, using the Gemini near-infrared spectrograph (GNIRS; Elias et al. 2006).  GNIRS was used in cross-dispersed mode  with the 111 l/mm grating, the short camera and the 0$\farcs$675 slit, giving   $R = 2800$. A central wavelength of 1.56 ~$\mu$m resulted in wavelength  coverage for orders 3 to 6 of 1.985 -- 2.175 ~$\mu$m, 1.589 -- 1.631    ~$\mu$m, 1.191 -- 1.305  ~$\mu$m and 0.993 -- 1.087  ~$\mu$m. This setting nicely sampled the flux peaks (Figure 2).

A total of 75  300$\,$s frames were obtained over five nights: 2014 March 17 and 22, 2014 May 18, 2014 July 11 and 13. Data were typically taken through thin clouds with 0$\farcs$7 seeing. An ``ABBA'' offset pattern was used with offsets of $3\arcsec$ along the slit.
The Gemini IRAF routines are not designed for this higher resolution cross-dispersed mode and so reduction was carried out manually. AB image pairs were subtracted from each other to form an image with a positive and negative spectrum. Pattern noise artefacts were then removed using a {\tt python} script designed for the purpose. Each image resulting from a subtracted pair was visually inspected, and some removed because of guider issues or low signal when at high airmass. To form the final coadded image, a total of 36 images with positive and negative spectral traces were combined, for a total of six hours on target. The IRAF {\tt apall} routine was used to extract spectra for each order, using the locations of the standard star spectral orders as a reference. One-dimensional multi-order spectra were extracted from the flat field and arc images obtained from the calibration lamp on the telescope using the locations of the science apertures as references. The one-dimensional multi-order spectra for science and standards were divided by the corresponding one-dimensional flat field spectrum and the positive and negative spectra for each order were then averaged, after multiplying the negative spectra by $-1$. The uncertainty in the spectrum was estimated from the difference between the positive and negative spectra. A wavelength solution was determined graphically from the arc spectrum and applied.

The standard stars observed to remove telluric absorption features and flux calibrate the spectra were HD 149803, an F7V star, and
HD 173494, an F6V star.  Template spectra for these spectral types were obtained from the spectral library of Rayner et al. (2009), and used to determine a one-dimensional sensitivity function. The sensitivity spectrum was then used to correct the shape of the target spectrum, in each order. 

Cushing et al. (2011) present  an $HST$ Wide Field Camera 3  G141 grism spectrum of W1738. We flux calibrated this spectrum using our revised $J$ and $H$ photometry, and scaled our higher resolution spectrum to match. We could not calibrate our spectrum directly because the spectral orders do not completely span the filter bandpasses. Figure 2 shows both our Gemini spectrum and the Cushing et al. $HST$  spectrum. Our spectrum has been smoothed with a 9 pixel boxcar. The uncertainty in the spectrum is also shown; 
S/N across the $YJH$ band peaks is $\sim 10$, while across the $K$ band it is  $\sim 5$.
 We compare our spectrum to models in \S 3.3.

\section{Comparison of Models and Data}

\subsection{Models}

In this work we use S12 and T15 cloud-free models only. 
Although chloride and sulphide clouds are important for T dwarfs with $T_{\rm eff}$ as low as 600~K  (Morley et al. 2012),  our focus is the 400~K Y dwarfs and it has been shown that cloud-free atmospheres reproduce observations better than existing cloudy models for such objects (L15). We also use
T15 models without the ad hoc modifications to the pressure-temperature gradient they considered, although such modifications can improve the agreement with near-infrared data for mid-type T dwarfs. The modifications were motivated by the possibility of atmospheric fingering convection induced by species condensation, which is ignored here. 

The known Y dwarfs are of necessity solar neighborhood objects, due to their instrinsic faintness. 
The majority of M dwarf stars in the local neighborhood have near-solar metallicity and age 
(e.g. Burgasser et al. 2015, Terrien et al. 2015) and the same is likely to be true of the Y dwarfs.
We restrict the models to surface gravities given by log $g =$ 4.0,  4.5 and 4.8 because evolutionary models show that these values correspond to an age range of 0.4 to 10 Gyr at $T_{\rm eff} \approx 400$~K (Saumon \& Marley 2008).  The  corresponding mass range is around 5 to 20 Jupiter masses, based on the evolutionary models.

Our analysis also uses T15 models whose nitrogen
and carbon chemistry is driven out of equilibrium by vertical mixing which
is parametrized with an eddy diffusion coefficient $K_{\rm zz}\,$cm$^2\,$s$^{-1}$.
The departure of the nitrogen chemistry from  equilibrium abundances is quite insensitive at these temperatures to the value of   $K_{\rm zz}$, however the carbon chemistry remains sensitive to mixing (Zahnle \& Marley 2014). The T15 models show that, at   $T_{\rm eff} = 400$~K,  models with $K_{\rm zz} = 10^8$ \,cm$^2$\,s$^{-1}$ produce a $\sim 30$\% stronger  CO absorption at $\lambda \approx 4.7 ~\mu$m than those with $K_{\rm zz} = 10^6$\,cm$^2$\,s$^{-1}$, as CO is enhanced at the expense of CH$_4$. We used the well-studied   $T_{\rm eff} = 600$~K T dwarf  ULAS J003402.77$-$005206.7 (Warren et al. 2007)  to constrain the value of the mixing coefficient. This object is the coolest brown dwarf with a {\em Spitzer} mid-infrared spectrum and, combined with other data and an accurate trigonometric parallax, its properties are well determined (Leggett et al. 2009, Smart et al. 2010).   We find that the T15 models with appropriate  $T_{\rm eff}$ and log $g$ (550 -- 600 K, 4.5 dex) give a [4.5] magnitude that is too faint by 0.4 magnitudes if $K_{\rm zz} = 10^8$ \,cm$^2$\,s$^{-1}$, but is within 0.1 magnitudes of the observed value if  $K_{\rm zz} = 10^6$ \,cm$^2$\,s$^{-1}$. Mixing may vary from object to object, and depend on  $T_{\rm eff}$,  log $g$ or metallicity. Previous studies have determined $K_{\rm zz}$ values of $10^2$ -- $10^6$\,cm$^2$\,s$^{-1}$  for L and T dwarfs (Geballe et al. 2009; Leggett et al. 2007, 2008, 2010;  Stephens et al. 2009) and for Jupiter $K_{\rm zz} \approx 10^8$\,cm$^2$\,s$^{-1}$ (Lewis \& Fegley 1983). Hence a value of  $K_{\rm zz} = 10^6$ \,cm$^2$\,s$^{-1}$ is reasonable for Y dwarfs and we adopt that value here.

The T15 models used here contain some improvements over those described in the T15 publication. The thermochemical data for H$_2$ were updated to be compatible with the JANAF database (Chase 1998) and the Saumon, Chabrier \& van Horn (1995) equation of state. For the disequilibrium models, the pressure-temperature profile was re-converged more frequently to ensure that the total flux is consistent with the effective temperature. Finally, the post-processed low-resolution spectra were computed with the correlated--$k$
method and the  high-resolution spectra  were computed using line-by-line opacities at a resolution of at least 1 cm$^{-1}$ (a higher resolution than used by T15, see also Amundsen et al. 2014). 

Apart from inclusion of non-equilibrium chemistry, the S12 and T15 models differ due to the inclusion by T15 of an updated CH$_4$ line list (Yurchenko \& Tennyson 2014), and the omission by T15 of PH$_3$. Also, although condensation
is included in the T15 models, the removal of the condensed species from the local gas is not. At these temperatures rain-out species  are not important opacity sources, however their inclusion in the gas may change the atmospheric opacities. 
For example not removing the condensed Fe allows it to react with H$_2$S to form FeS, removing the   H$_2$S absorption (Morley et al. 2014, Marley \& Robinson 2015). 
The two model sets also use different solar abundances (Lodders 2003 for S12, Caffau et al. 2011 for T15), and different treatments of line broadening. 

Figure 3 compares  $T_{\rm eff} = $ 400~K and log $g = $ 4.0 solar metallicity models from S12 and T15. The top panel compares S12 and T15 equilibrium chemistry models, and the lower panel compares T15 equilibrium and non-equilibrium models. Filter bandpasses are shown for reference. The reader is referred to L15 (their Figure 5), and Figures shown later in this paper, for identification of the species causing the pronounced absorption bands in these spectra at this temperature. The dominant opacity sources are H$_2$,  H$_2$O, CH$_4$, and NH$_3$; CO and PH$_3$ may be important at $\lambda \sim 5 ~\mu$m. 

Comparison of the S12 and T15 spectra for equilibrium chemistry and $T_{\rm eff} =$ 400~K (Figure 3 top panel), shows that the S12 spectrum is brighter at $0.99 < \lambda ~\mu$m~$  < 1.33$, $1.98 < \lambda ~\mu$m~$ < 2.58$ and  $8.9 < \lambda ~\mu$m~$ < 9.9$
but fainter at  $4.10 < \lambda ~\mu$m~$ < 4.62$, by factors up to 1.5 --- 2.0; at other wavelengths the spectra are very similar. We suspect that the difference at $\lambda \sim 4.3 ~\mu$m is due to the omission of PH$_3$ by T15 (see L15, Figure 5, middle panel). The differences at other wavelengths may be due to the use of the new CH$_4$ line list by T15. At these low temperatures the near-infrared spectrum is very sensitive to opacity changes, as demonstrated by the changes seen when the new H$_2$ and NH$_3$ opacities were incorporated into the S12 models (see S12 Figure 7, bottom panel). 

Comparison of T15 spectra with and without non-equilibrium chemistry for  $T_{\rm eff} = 400$ K  (Figure 3 bottom panel) shows that the $H$ band is much brighter in the  non-equilibrium case, as are the $ 3 ~\mu$m and $10 ~\mu$m regions. This is because the abundances of CH$_4$ and NH$_3$ are reduced in favor of CO and N$_2$. Because CO is enhanced, the  $ 4.7 ~\mu$m region is fainter  in the  non-equilibrium case.  The blue wings of the $Y$ and $K$ bands are brighter because of the reduction in NH$_3$. The cause of the decrease in flux near the peak of the $K$ band in the non-equilibrium case is not clear -- there are no opacities in this region that should be enhanced by mixing. It may be that the changes introduced into the atmosphere, by mixing, 
redistributes the flux to regions previously impacted by CH$_4$ or NH$_3$ absorption. Although not easily seen in Figure 3, there is a similar decrease in flux near the $Y$ band peak in the non-equilibrium model.  We discuss this further in Section 3.3.

We compare the models to data in the following sections.

\subsection{Models and Photometry}

Figures 4 and 5 present near-infrared color-magnitude and color-color diagrams for late-T and Y dwarfs, where observational data are compared to cloud-free solar metallicity S12 and T15 model sequences. Data sources are this work and L15, as well as earlier publications referenced in L15. Brown dwarfs with $M_J > 19$ are labelled in Figure 4, as well as known binary systems where one component has  $M_J > 19$; the Appendix gives full names and discovery references for these objects. Filter bandpasses are shown in Figure 3 and in Figure 6 below.

Figure 4 shows  $M_J$ as a function of $J - H$ (left) and $J -$ [4.5] (right). S12 and T15 equilibrium sequences are shown for log $g =$ 4.5, and T15 non-equilibrium models are shown for  log $g =$ 4.0,  4.5 and  4.8, as identified in the legends. Comparing the T15 equilibrium and non-equilibrium models in each panel shows that the brightening of the $H$ band and the reduction in the  [4.5] flux in the non-equilibrium models (see Figure 3) improves the agreement with the observations, particularly for  $T_{\rm eff} > 450$ K. 
   
Figure 5 shows various color-color plots where observations are compared to T15  log $g =$ 4.0, 4.5 and  4.8
non-equilibrium model sequences, and the T15  log $g =$ 4.5 equilibrium sequence. 
As previously mentioned, the agreement at $J - H$ is much improved with the inclusion of non-equilibrium chemistry, and the agreement at [3.6] $-$ [4.5] is also improved for the T dwarfs with  $T_{\rm eff} > 500$ K. The $Y - J$ and $J - K$ colors do not support one treatment of the chemistry over the other, at least for the current models. There are large discrepancies in the $J - K$ and [3.6] $-$  [4.5] colors which get worse for redder $J -$ [4.5], or lower temperatures. Both $K$ and [3.6] are too faint in the T15 models (also in the S12 models, see L15). 
Very little flux is emitted at $2 < \lambda ~\mu$m $< 4$ (Figure 3) and so this inadequacy in the models is not expected to significantly impact the atmospheric structure, but it is desirable to resolve the issue.   In this wavelength region both NH$_3$ and CH$_4$ are important opacity sources (e.g. L15), and it is possible that adjusting the carbon and nitrogen mixing can address the problem. The discrepancies are discussed further below, where we compare the models to near-infrared spectra.

Figure 5 indicates that the modeled $Y - J$ and $J - K$ colors are sensitive to gravity for $J -$ [4.5] $>$ 5, or  $T_{\rm eff} < 450$ K (Figure 4). To explore this further, Figure 6 shows three T15 synthetic near-infrared spectra for  $T_{\rm eff} = 400$ K  and  $K_{\rm zz} = 10^6$\,cm$^2$\,s$^{-1}$. The models differ either in gravity or metallicity, as indicated by the legend.  We calculated a very small number of non-solar-metallicity models, motivated by the  observed dispersion in the observational data in Figures 4 and 5,
and the knowledge  that metallicity does impact the spectral energy distribution of brown dwarfs (see e.g.  Burningham et al. 2013). A more complete study of the effect of metallicity will be done elsewhere. 

Figure 6 shows that a decrease in gravity or an increase in metallicity increases the flux emerging at $K$. 
This is a well-known effect which is due to the relative importance of pressure-induced
H$_2$ opacity which increases with gravity and decreases with metallicity (e.g. Borysow et al. 1997, S12).
Based on these models, a change in metallicity of 0.2 dex has a much larger impact on the $K$ magnitude than a change in gravity of 0.5 dex. Figure 6 also shows that the flux emerging from the blue half of the $Y$ band is sensitive to gravity and metallicity. Here the changes go in the opposite sense, so that an increase in metallicity or decrease in gravity reduces the emerging flux. Also, the gravity change of 0.5 dex has a larger impact than the metallicity change of 0.2 dex on the  $Y$ magnitude, according to these models. 
The primary opacity source at $\lambda \approx 1\,\mu$m is H$_2$O, which, based on the small changes seen in  the wings of the $J$ flux peak, does not appear to be sensitive to  these parameters. We will explore nature of the $Y$ band sensitivity to gravity in future work.

\subsection{Models and Spectroscopy}

\subsubsection{Comparison Method}

In this section we compare near-infrared spectra of three Y dwarfs to T15   synthetic spectra. The model to observation comparison is done by eye, and the $K$ band is neglected because of the inadequacies in the models in this region (Figure 5). 
The Y dwarfs are the Y0--0.5 dwarf  WISEPC J121756.91+162640.2B (W1217B; Kirkpatrick et al. 2012, Liu et al. 2012), the Y0 dwarf W1738 and  the Y1 dwarf W0350.
We use the  $T_{\rm eff}$ values indicated by the  $J -$ [4.5] color (Figure 4) as the starting point for spectral fits, as this mimics the spectral shape from 1 ~$\mu$m to 5 ~$\mu$m. Most of the energy is emitted in the  5 ~$\mu$m window for the late-T and Y dwarfs (e.g. Morley et al. 2012, 2014), and therefore it is important to include this region in any temperature estimate. Figure 4 shows that we expect $T_{\rm eff}$ values of  400 -- 450 K for W1217B and W1738, and 325 -- 375 K for W0350. 
The model spectra used in the comparison are either in chemical equilibrium or have log $K_{\rm zz} = 6$; have surface gravities given by
log $g =$ 4.0, 4.5 and 4.8; and have $T_{\rm eff}$ (K) $=$  325, 350, 375, 400, 425 and 450.

For each Y dwarf, then, we have model spectra with three values of  $T_{\rm eff}$, and for each of these we have three values of log $g$. We compared the nine model spectra to the observed near-infrared spectrum, where we scale the flux of the model spectra (generated for the Y dwarf surface) by the square of the Y dwarf radius and the inverse-square of the distance to the Y dwarf. The Y dwarf radius for any  $T_{\rm eff}$ and $g$ combination is obtained from Saumon \& Marley (2008) evolutionary models. The distances to W1217B and W1738 are taken from published values of trigonometric parallax, although we allowed adjustments in the overall brightness of the model spectra corresponding to the $1 \sigma$ quoted uncertainties in the parallax measurements.   For W0350 only a preliminary parallax measurement is available, and we find that a very large adjustment is needed, outside of the quoted uncertainties; we discuss this in \S 3.3.4.  As mentioned previously, photometric variability at the $\sim 10$\% level is possible (Crossfield 2014). We neglect any change in near-infrared spectral shape due to variability,  assuming that that any wavelength-dependence is not significant for the cloud-free Y dwarfs.

\subsubsection{W1217B}

Figure 7 shows the spectrum of the Y0--0.5 dwarf  W1217B presented by Leggett et al. (2014). 
Principal absorbers are indicated;  for more detailed identifications of features in near-infrared spectra of cool brown dwarfs the reader is referred to Bochanski et al. (2011) and Canty et al. (2015).
Leggett et al. studied the properties of the W1217AB system using coevality and luminosity arguments for the binary, as well as comparisons of the photometry and spectroscopy of the components to Morley et al. (2012) equilibrium models.
The authors determined $T_{\rm eff} =$ 450~K, log $g =$ 4.8, a possibly sub-solar metallicity, and thin to no clouds.
T15 fit the same observed spectrum using a model with $T_{\rm eff} =$ 425~K, log $g =$ 4.0 and  log $K_{\rm zz} = 8$. Once the evolutionary radius is used the T15-selected model spectrum would be too bright, however that could be compensated by adjusting the distance to the brown dwarf, which is quite uncertain. 

We find that the current suite of T15 models shows that the best fit is given by the $T_{\rm eff} =$ 450~K, log $g =$ 4.5 and  log $K_{\rm zz} = 6$ model spectrum, although this requires a distance of 11.3 pc, compared to the measured value of $10.1^{+1.9}_{-1.4}$ pc (Dupuy \& Kraus 2013).  The second-best fit is the $T_{\rm eff} =$ 450~K, log $g =$ 4.8 and  log $K_{\rm zz} = 6$ model (Leggett et al. 2014 also find $T_{\rm eff} =$ 450~K and log $g =$ 4.8). This model spectrum would require no adjustment of the distance, however the fit is poorer at $Y$ and $K$, and similar at $J$ and $H$, as shown in Figure 7. 

Figure 7 also shows the best fitting equilibrium chemistry model, and demonstrates the impact of mixing in the near-infrared more clearly than Figure 3. Including mixing greatly improves the fit in the region of the strong   NH$_3$ absorption at 
$\lambda \approx 1.03 ~\mu$m and $\lambda \approx 1.55 ~\mu$m. However the equilibrium model better reproduces the data at  $1.055 \lesssim \lambda\,\mu$m $ \lesssim 1.09$
and  $2.07 \lesssim \lambda\,\mu$m $ \lesssim 2.15$. 

Overall,  the selected  $T_{\rm eff} =$ 450~K, log $g =$ 4.5 and  log $K_{\rm zz} = 6$ 
spectrum gives a superior fit to the relative heights of the $Y$, $J$ and $H$ flux peaks compared to the  previously published fits by
T15 and Leggett et al. (2014).

\subsubsection{W1738}

Figure 8 shows our best fit to the new $R = 2800$ spectrum presented here for the Y0 dwarf W1738.  
The best fit in this case is provided by the $T_{\rm eff} =$ 425~K, log $g =$ 4.0 and  log $K_{\rm zz} = 6$ model spectrum, with no adjustment needed to the measured distance of $7.8 \pm 0.6$ pc (Beichmann et al. 2014).

Motivated by the low model flux at $K$, we compared the observations to a small number of 
super-solar metallicity models, and found that a model with [m/H] $= +0.2$, $T_{\rm eff} =$ 400~K, log $g =$ 4.0 and  log $K_{\rm zz} = 6$ gives almost as good a fit, however the shape of the $Y$ flux peak is poorer, as shown in Figure 8. This slightly cooler, higher metallicity model matches the data better if the distance is reduced to the low end of the  range measured by Beichmann et al.. 

Figure 8  also compares the observations to the best fitting equilibrium chemistry model
with  $T_{\rm eff} =$ 450~K and log $g =$ 4.5. In this case the match is improved if the distance in increased towards the high end of the measured range. The expected equilibrium-chemistry problem of overly strong NH$_3$ absorption  $\lambda \approx 1.03 ~\mu$m and $\lambda \approx 1.55 ~\mu$m is seen, and also the relative $YJH$ flux peaks are not as well-reproduced by this model.

\subsubsection{W0350}

Figure 9 shows our best fit to the new $R = 540$ spectrum presented here for the Y1 dwarf W0350.  The apparent drop in observed flux at $1.06 \lesssim \lambda ~\mu$m $\lesssim 1.07$ should be confirmed by new observations (flux calibration of spectra at the extremes of the wavelength range can be prone to error due to the rapidly changing instrument sensitivity function). If real, it may be an indicator of water clouds (e.g. Morley et al. 2014, their Figure 10), or it may provide an additional constraint on gravity or metallicity (Figure 6). 

For this brown dwarf, the preliminary trigonometric parallax published by Marsh et al. (2013) implies an unrealistically faint absolute magnitude of $M_{[4.5]} = 16.9$, which would make W0350 similar in luminosity to the extreme dwarf W0855 while $J -$ [4.5] is more than 3.5 magnitudes bluer (see Figure 4). In matching the spectra, we start with the model set constrained  in temperature and gravity as described in \S 3.1 and \S 3.3.1, and radii given by evolutionary models, but allow the distance to be greater than implied by the parallax measurement of $3.4^{+0.7}_{-0.5}$ pc. An error in the preliminary Marsh et al. results, especially in the harder to measure smaller-parallax greater-distance direction, would not be surprising.  For comparison, Beichman et al. (2014) revise the distance for another Y dwarf in their sample (WISE J041022.71+150248.4) from $4.1^{+1.6}_{-0.2}$ to  $6.2 \pm 0.4$.

The best fit is provided by the $T_{\rm eff} =$ 350~K, log $g =$ 4.0 and  log $K_{\rm zz} = 6$ model spectrum.
At a higher gravity of log $g =$ 4.5 the fit is almost as good, but the $H$ flux peak appears to have an excess of flux at $\lambda > 1.59 ~\mu$m, as shown in Figure 9.  
The fits imply a distance of 6.3 pc for the preferred model, and 5.3 pc for the second model.   In the absolute magnitude diagrams in Figure 4 we use our distance of  6.3 pc. 

Figure 9 also shows the best fitting equilibrium chemistry model
 with  $T_{\rm eff} =$ 350~K and log $g =$ 4.5. This fit implies a distance of 5.3 pc, as found for the non-equilibrium model with the same values of  $T_{\rm eff}$  and $g$.
There is again evidence that the equilibrium model is too faint in the blue wing of the $H$-band, suggesting that even at the low temperature of  $T_{\rm eff} =$ 350~K mixing of N$_2$ and NH$_3$ is significant. 

\subsubsection{Quality of Fit}

Although the non-equilibrium chemistry model fits to the spectra of the three Y dwarfs are remarkably good in some regions, especially the $J$ band, systematic offsets can be seen in Figures 7, 8 and 9. In the $Y$ band,  the model flux at $1.01 \lesssim \lambda ~\mu$m $\lesssim 1.04$ is low, as is the flux at  $1.06 \lesssim \lambda ~\mu$m $\lesssim 1.09$.   In the $H$ band,  the model flux at $1.51 \lesssim \lambda ~\mu$m $\lesssim 1.555$ is slightly low, and the model flux at  $1.57 \lesssim \lambda ~\mu$m $\lesssim 1.60$ is high. In particular, there is a strong absorption feature observed at $\lambda \approx 1.59 ~\mu$m that is not seen in the models (Figures 7, 8, 9). In the $K$ band  the model flux at $2.02 \lesssim \lambda ~\mu$m $\lesssim 2.18$ is low by a factor of 2 -- 3. The photometric comparison presented above indicates that the model flux is also too low at  $3 < \lambda ~\mu$m $ < 4$.

Most of these problem regions can be identified as areas where NH$_3$ and CH$_4$ absorption is important. The strong absorption at  $\lambda \approx 1.59 ~\mu$m looks similar to the feature seen in the spectrum of the 500~K brown dwarf UGPS J072227.51−054031.2, and identified by Canty et al. (2015) as a combination of both  NH$_3$ and CH$_4$ absorption. Figures 3, 7 and 8 show that there is a decrease in flux at the $Y$ and $K$ flux peaks when mixing is included in the models, the cause of which may be simple flux redistribution (\S 3.1). This effect appears to be too strong, based on our comparisons.  All of these issues suggest that  tuning the chemical mixing, or using a  retrieval approach such as  in Line et al. (2015), would improve the models.

Using the preferred model parameters for W1738 and W0350 (resolved mid-infrared photometry does not exist for the W1217AB system), the T15 $J -$ [4.5] colors for these two objects are about 0.4 magnitudes larger than observed (Figure 4). 
Considering the uncertainties in the fits and the remaining deficiences in the models, the agreement is reasonable.
Table 1 gives the values of  $T_{\rm eff}$ and log $g$ for W0350 and W1738 based on the model fits to the new spectra presented here, with the corresponding mass and age from evolutionary models. Note that the lower gravity we find for W1217B compared to Leggett et al. (2014) implies a younger age for the system of 1 -- 5 Gyr and a lower mass for W1217B of 10 -- 15 M$_{\rm Jupiter}$, compared to the 4 -- 8 Gyr and 20 -- 24  M$_{\rm Jupiter}$ found by those authors.


\section{Estimated Physical Parameters for W0350 and W1738}

Evolutionary models show that mass is tightly constrained by the surface gravity for solar-neighborhood objects at these temperatures (see e.g. Saumon \& Marley 2008, their Figure 4, and Allard et al. 1996, Marley et al. 1996, Burrows et al. 1997).  Changing log $g$ by 0.5 dex changes the [4.5] flux by  a factor $\approx 1.25$, which is larger than or similar to the likely uncertainty in the model. However our experiments using the $YJH$ peaks together in a spectral by-eye comparison shows that gravity can be further constrained to 0.25 dex for these Y dwarfs (e.g. Figure 7). The fits, together with evolutionary models, then imply a mass of $5^{+4}_{-2}$
Jupiter masses for both W0350 and W1738. Once the models are more robust, the $Y$ and $K$ colors show promise for constraining log $g$ (Figures 5 and 6),  although metallicity variations  will need to also be considered.

Given the range in colors seen for different gravities and different input physics (S12 vs. T15) in
Figure 4, the uncertainty in our derived temperature is $< 50$~K. Current models show that a change in 25~K at these temperatures changes the flux in the $J$ band  by a factor $\gtrsim 1.5$ which is larger than the likely systematic error in the models and data, and we adopt an uncertainty in  $T_{\rm eff}$ of $\sim \pm 25$~K. This is supported by the range in parameters for the preferred and runner up models described in the previous section. Note that exploration of a limited set of non-solar metallicity models in the previous section showed that a change in [m/H] of 0.2 dex changed the derived  $T_{\rm eff}$ by 25~K, consistent with the adopted uncertainty (Figure 9).

Gravity and temperature together constrain age (see e.g. Saumon \& Marley 2008, their Figure 4). Taking into account uncertainties of 25~K and 0.25 dex in  $T_{\rm eff}$ and log $g$ respectively, W0350 has an age of  0.3 -- 3 Gyr and W1738 has an age of 
0.15 -- 1 Gyr. Interestingly, W0350 seems to be a cooler, older version of W1738, as both have a mass around 5 Jupiter masses. 
Both W0350 and W1738 have low tangential velocities of around 20 km s$^{-1}$, based on the proper motions of Marsh et al. (2013) and Beichman et al. (2014), and the distance used here for W0350 and that determined by Beichman et al. for W1738. This is consistent with thin disk kinematics (e.g. Dupuy \& Liu 2012) and an age $<$ 7 Gyr (e.g. Brook et al. 2012), consistent with the 
loose constraints on age determined here.

\section{Conclusion}

As expected, models which include vertical mixing and the resulting non-equilibrium chemistry reproduce observations of Y dwarfs better than those which do not include mixing. Hydrogen, carbon, nitrogen and oxygen chemistry shapes the spectral energy distributions of these cold brown dwarfs. Vertical mixing in the atmosphere impacts the chemistry of carbon and nitrogen, and the remaining systematic discrepancies between the data and models that we have identified, where both NH$_3$ and CH$_4$ are important, could be addressed by fine-tuning the mixing. The kinetics of nitrogen species are uncertain, and Moses (2014) shows that different treatments of quenching do result in significantly different NH$_3$ abundances; this will be an avenue of future exploration. Mixing and non-equilibrium chemistry is also expected to increase the abundance of PH$_3$ and HCN in cool atmospheres (Zahnle \& Marley 2014, Sousa-Silva et al. 2015), which has not been taken into account in current models, and which should be included in future models.

The model comparisons show that for $T_{\rm eff} \approx 400$ K Y dwarfs, the atmospheric chemistry needs to change such that, while conserving flux:
\begin{itemize}
\item
the CH$_4$ absorption is decreased at $2 < \lambda ~\mu$m $< 4$ without increasing the CO absorption at $\lambda = 4.7 ~\mu$m
\item
the CH$_4$ and NH$_3$ absorption at $\lambda \approx 1.59 ~\mu$m is increased without increasing the absorption elsewhere in the near-infrared
\item
the NH$_3$ absorption at $\lambda \approx 1.03 ~\mu$m is decreased
\item
the flux at $\lambda \approx 1.08 ~\mu$m and  $\lambda \approx 2.10 ~\mu$m is increased without strengthening the CH$_4$ or NH$_3$ absorption, except at  $\lambda \approx 1.59 ~\mu$m
\end{itemize}
It may be easiest to achieve this by applying a retrieval analysis such as done recently for late-T dwarfs by Line et al. (2015). On the other hand, less than 10\% of the total flux is emitted at $\lambda < 3.5\,\mu$m
in these objects, all in the Wein tail of the Planck function. This makes the near-infrared
spectrum very sensitive to details of the opacity and chemical abundances. A more robust analysis
would be possible with spectroscopic data beyond $3\,\mu$m.

New or improved trigonometric parallaxes for the Y dwarfs would be valuable. Less uncertain distances would allow tighter constraints on gravity and temperature when fitting spectra as we did here, incorporating the brown dwarf radius and distance and requiring that the absolute flux levels be consistent.

Despite the remaining discrepancies and uncertainties, the quality of the models and the observational data are quite impressive, and the fits that we show here to $J$ band spectra in particular, are remarkably good. We find that the isolated Y dwarfs   WISE J035000.32$-$565830.2 and WISEP J173835.52$+$273258.9 both have a mass of  $5^{+4}_{-2}$ Jupiter masses, and their ages are 
0.3 -- 3 Gyr and 0.15 -- 1 Gyr respectively. This is consistent with the low tangential velocity of  around 20 km s$^{-1}$ measured for both brown dwarfs and puts them well within the
commonly accepted mass range of planets.

\acknowledgments

Based on observations obtained at the Gemini Observatory, which is operated by the Association of Universities for
Research in Astronomy, Inc., under a cooperative agreement with the NSF on behalf of the Gemini partnership: the National Science Foundation (United States), the Science and Technology Facilities Council (United Kingdom), the National Research Council (Canada), CONICYT (Chile), the Australian Research Council (Australia),
    Minist\'{e}rio da Ci\^{e}ncia, Tecnologia e Inova\c{c}\~{a}o (Brazil)
    and Ministerio de Ciencia, Tecnolog\'{i}a e Innovaci\'{o}n Productiva
    (Argentina). S. L.'s research is supported by Gemini Observatory.  D.S.' work was supported in part by 
NASA grant
      NNH12AT89I from Astrophysics Theory. I. B.'s work is supported by the European Research Council through grant ERC-AdG No. 320478-TOFU.This publication makes use of data products from the Wide-field Infrared Survey Explorer, which is a joint project of the University of California, Los Angeles, and the Jet Propulsion Laboratory/California Institute of Technology, funded by the National Aeronautics and Space Administration. This research has made use of the NASA/ IPAC Infrared Science Archive, which is operated by the Jet Propulsion Laboratory, California Institute of Technology, under contract with the National Aeronautics and Space Administration.

\appendix

\section{Brown Dwarf Identifications}

Table 2 lists the full name of the T and Y dwarfs  identified in Figure 4, together with discovery references.

\clearpage

\begin{figure}
    \includegraphics[angle=-90,width=.99\textwidth]{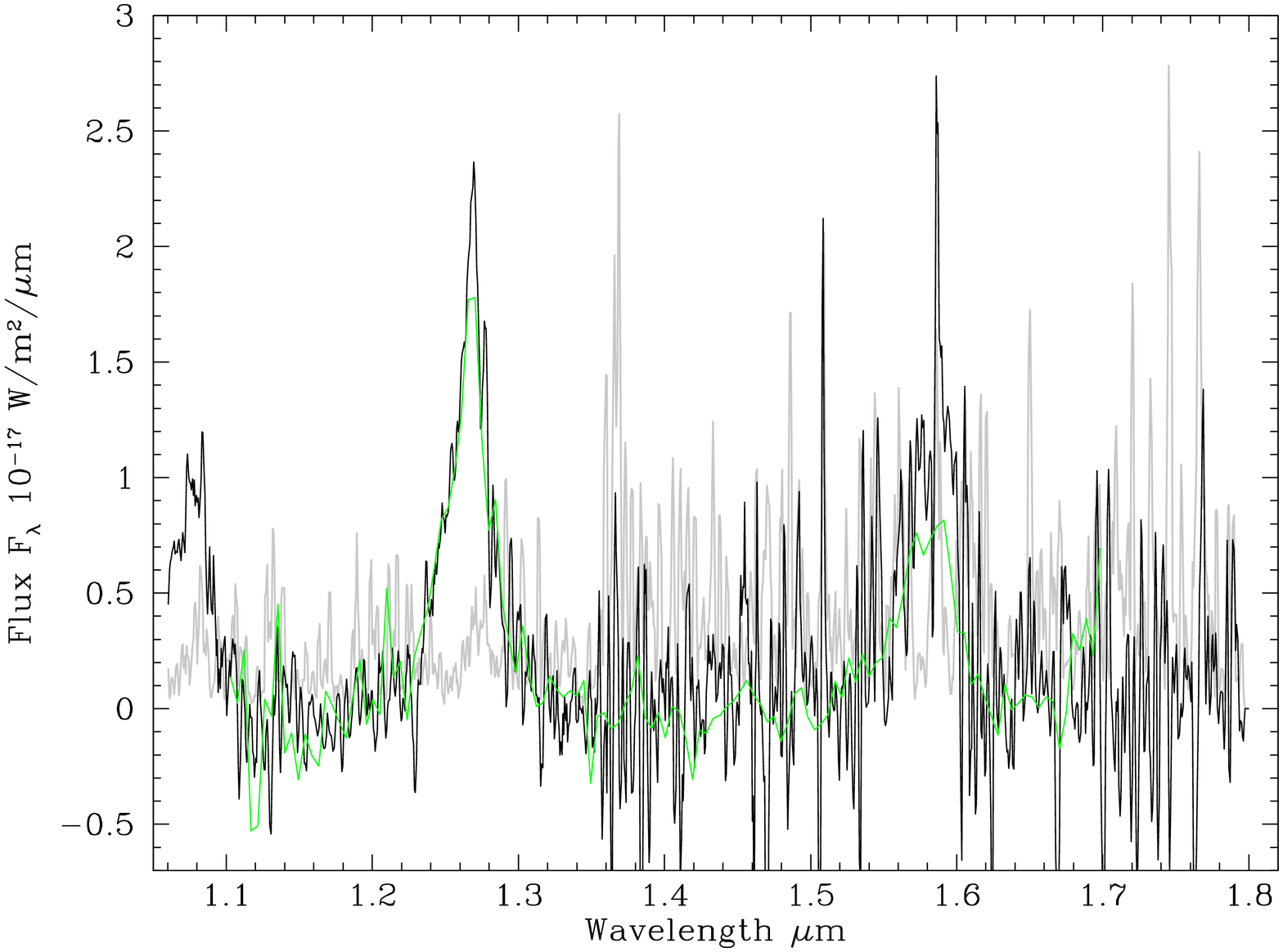}
\caption{The black line is the spectrum of WISE J035000.32$-$565830.2 obtained using Gemini and
presented in this work,
and the green line is a lower resolution spectrum obtained by Schneider et al. (2015) using {\it HST}. 
 The gray line is the uncertainty in the Gemini observational data.
\label{fig1}}
\end{figure}

\begin{figure}
    \includegraphics[angle=-90,width=.99\textwidth]{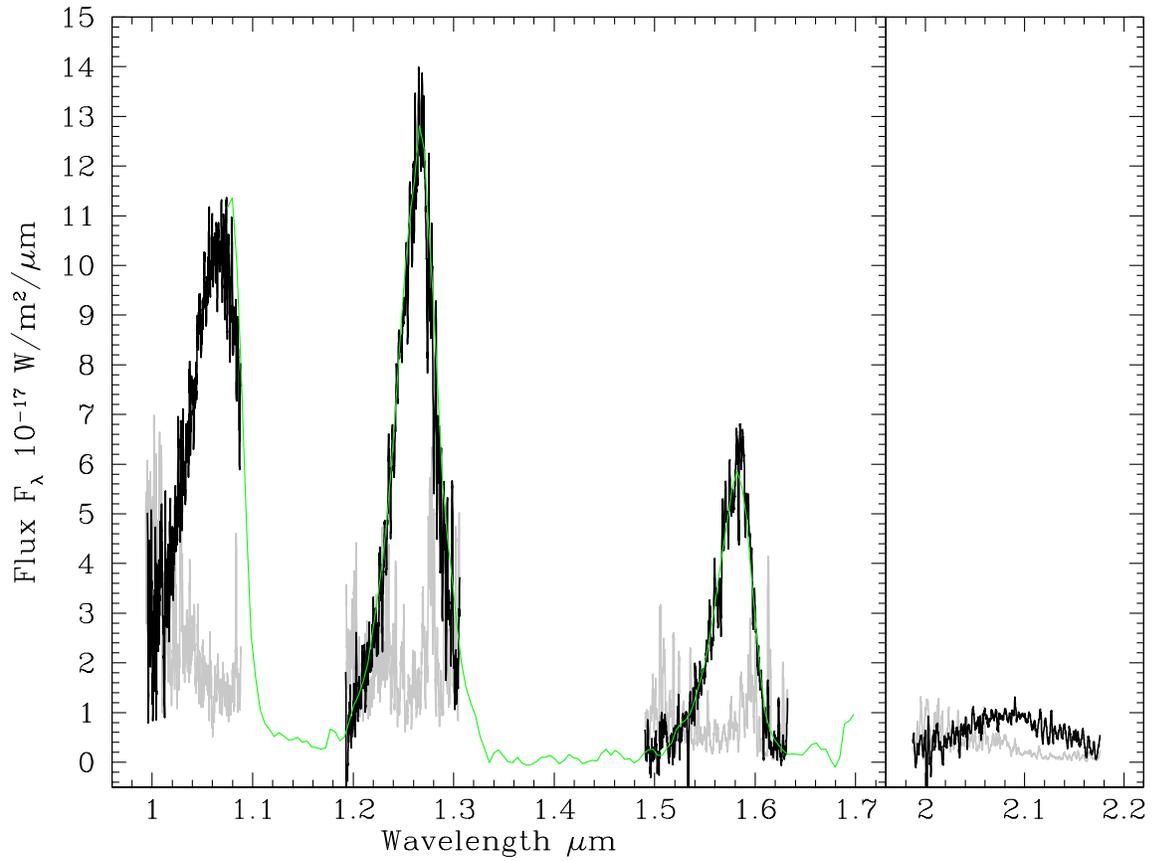}
\caption{The black line is the spectrum of WISEP J173835.52$+$273258.9  obtained  using Gemini and
presented in this work,
and the green line is a lower resolution spectrum obtained by Cushing et al. (2011) using {\it HST}. 
 The gray line is the uncertainty in the Gemini observational data.
\label{fig2}} 
\end{figure}

\begin{figure}
    \includegraphics[angle=-90,width=.99\textwidth]{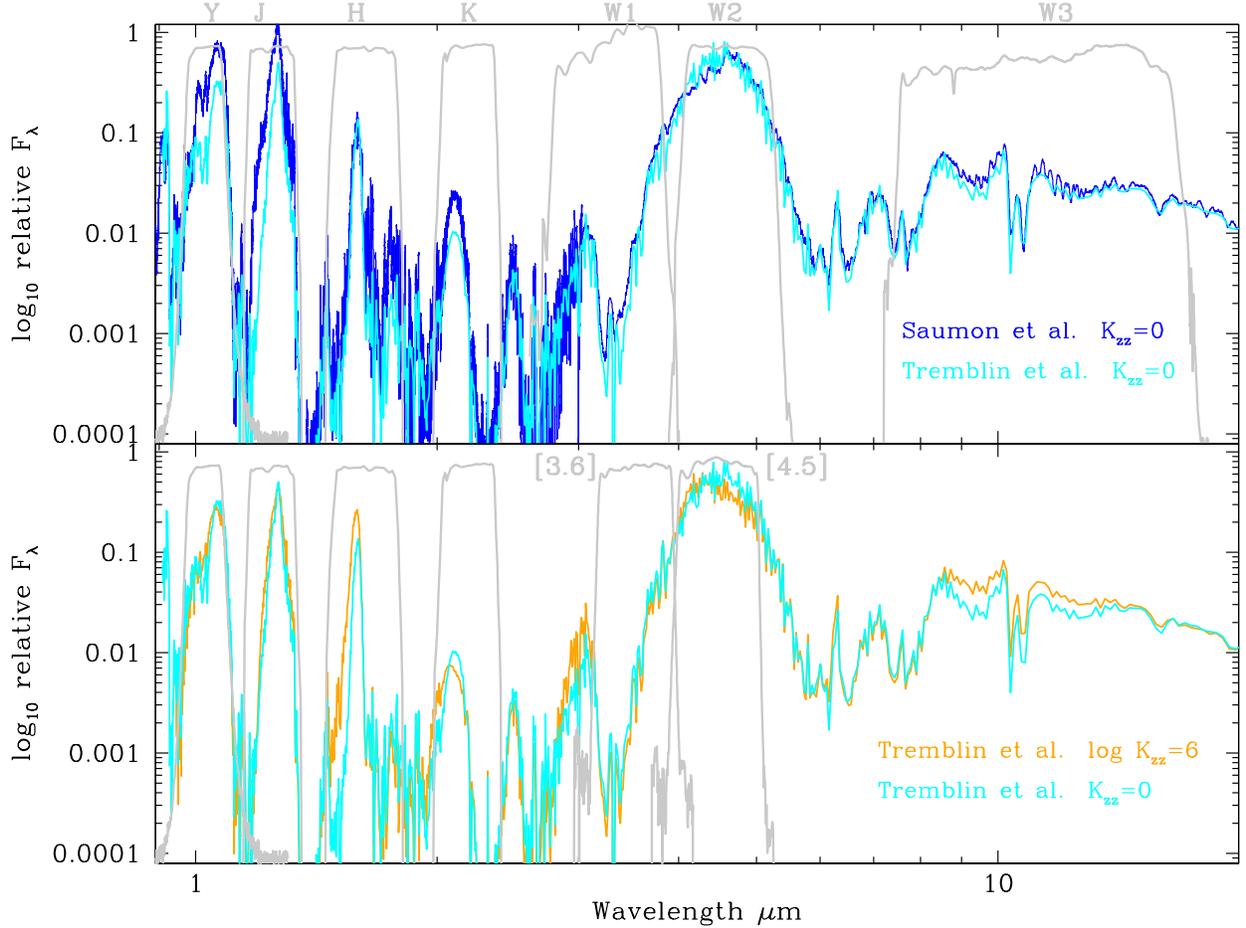}
\caption{
Comparison of Saumon et al. (2012) and Tremblin et al. (2015) cloud-free model spectra for $T_{\rm eff} = $400~K, log $g = $4.0, solar metallicity atmospheres. The spectra have been smoothed to $R \sim 1000$. Near-infrared MKO-system passbands are shown, as well as the mid-infrared $WISE$ filter passbands (upper panel) and IRAC $Spitzer$ passbands (lower panel). The upper panel compares equilibrium chemistry models and the lower panel compares equilibrium and non-equilibrium models. Model parameters are given in the legends. See text for discussion.
\label{fig3}}
\end{figure}

\begin{figure}
    \includegraphics[angle=-90,width=1.01\textwidth]{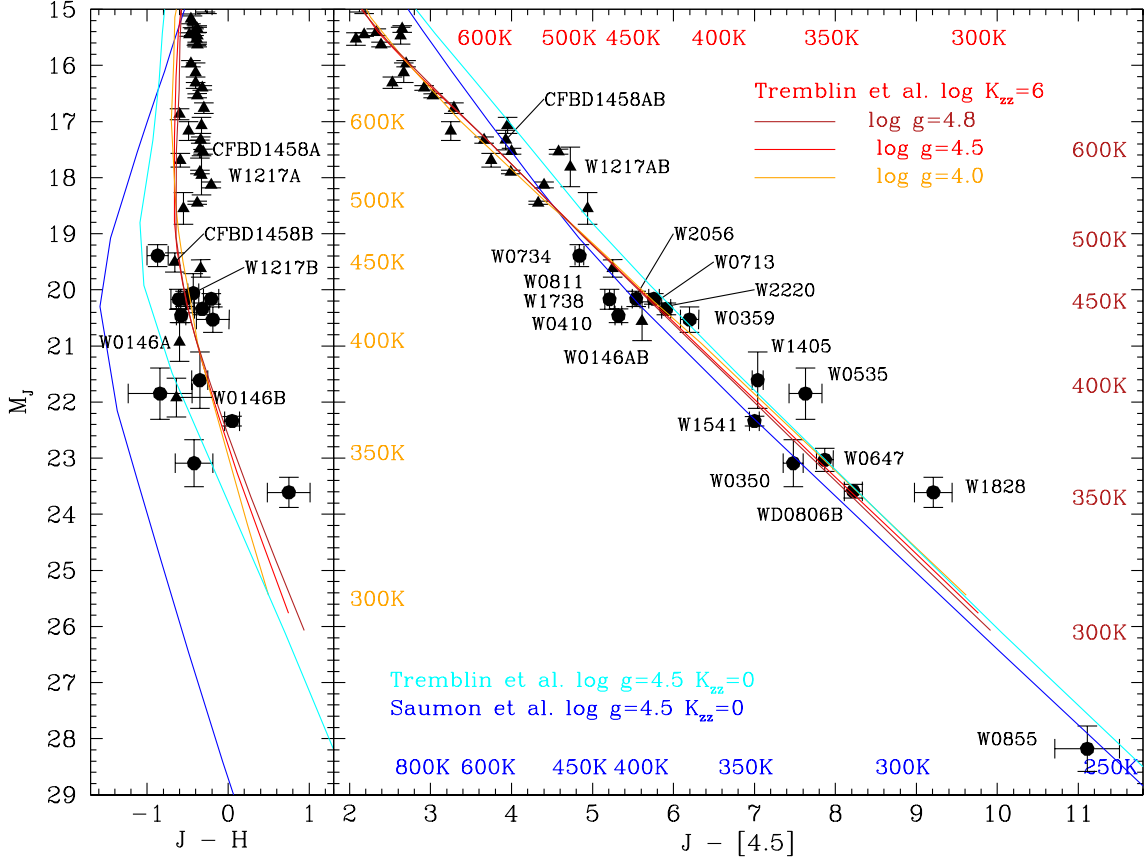}
\caption{$M_J$ as a function of $J - H$ and $J -$ [4.5]. S12 and T15 equilibrium sequences are shown for log $g =$ 4.5, and T15 non-equilibrium models are shown for  log $g =$ 4.0, 4.5 and 4.8, as indicated in the legends. Color-coded $T_{\rm eff}$ values for the models are shown along the axes. Data points are colors and magnitudes for late-T and Y dwarfs (triangles and circles respectively), using the MKO near-infrared system and the IRAC Vega-based system. Sources of photometry and parallax are this work and as referenced in L15.   
The Y dwarf W0350 is plotted using the photometric parallax determined here (see \S 3.3.4) and so $M_J$ agrees with the models by definition.
The close T/Y binary systems CFBD1458AB, W1217AB and W0146AB have resolved near-infrared photometry but not mid-infrared. 
Error bars along the $x$ axes are omitted for the T dwarfs, for clarity. Full names for the labelled sources are given in the Appendix.
\label{fig4}}
\end{figure}

\begin{figure}
    \includegraphics[angle=-90,width=1.05\textwidth]{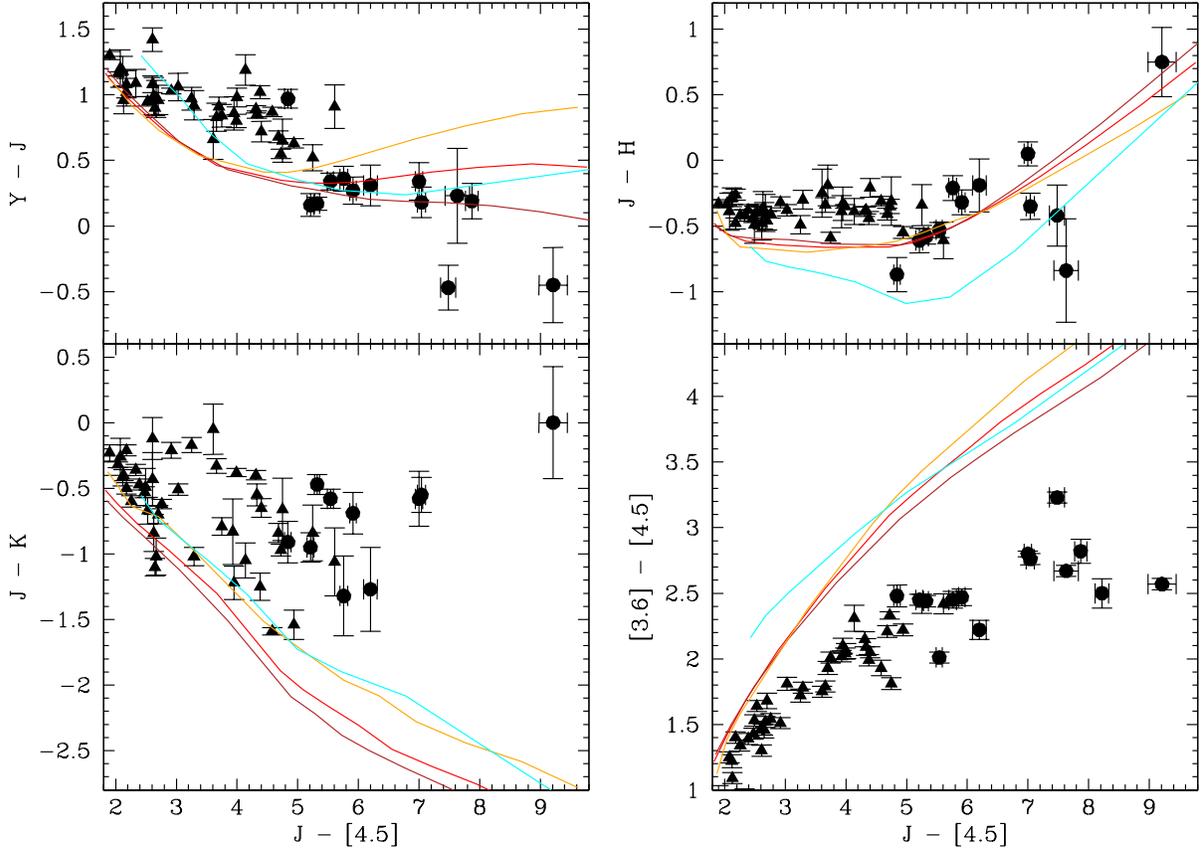}
\caption{Color-color plots comparing observations to T15  log $g =$ 4.0 (orange line), log $g =$ 4.5 (red line) and  log $g =$ 4.8 (dark red line)
non-equilibrium model sequences. A T15 equilibrium model sequence with log $g =$ 4.5 (cyan line) is also shown. Symbols  are as in Figure 4.  Not shown, for clarity, is the extreme dwarf  W0855; this dwarf has no published $Y$, $H$ or $K$ values, and $J -$ [4.5] $= 11.1 \pm 0.4$ and [3.6] $-$ [4.5] $= 3.55 \pm  0.05$. 
\label{fig5}}
\end{figure}

\begin{figure}
    \includegraphics[angle=-90,width=1.05\textwidth]{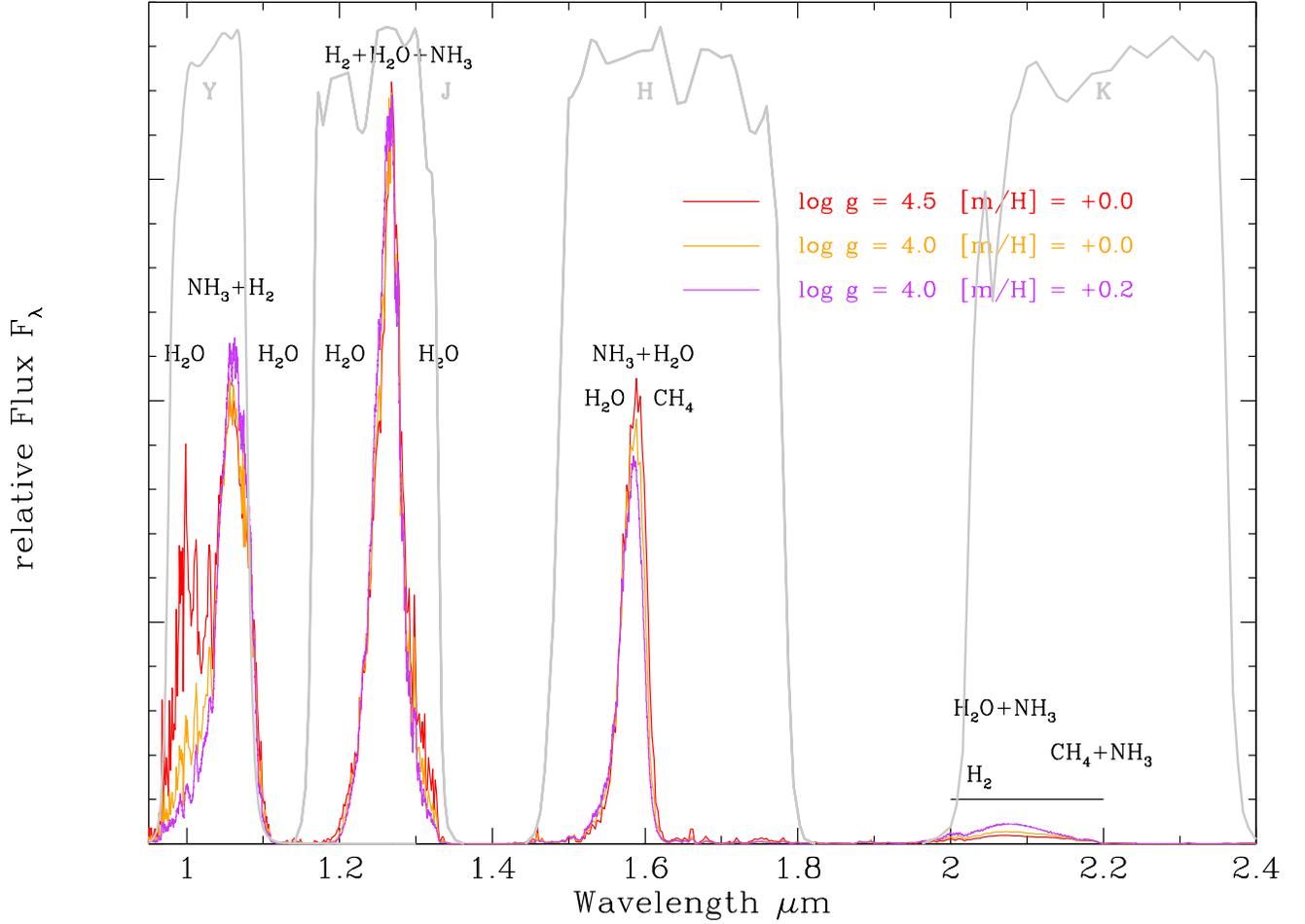}
\caption{Synthetic T15 near-infrared spectra for  $T_{\rm eff} = 400$ K  and  $K_{\rm zz} = 10^6$\,cm$^2$\,s$^{-1}$. The models differ either in gravity or metallicity, as indicated by the legend. The spectra have been scaled to match at the $J$ band peak. MKO near-infrared filter profiles are shown.
\label{fig6}}
\end{figure}

\begin{figure}
    \includegraphics[angle=-90,width=1.02\textwidth]{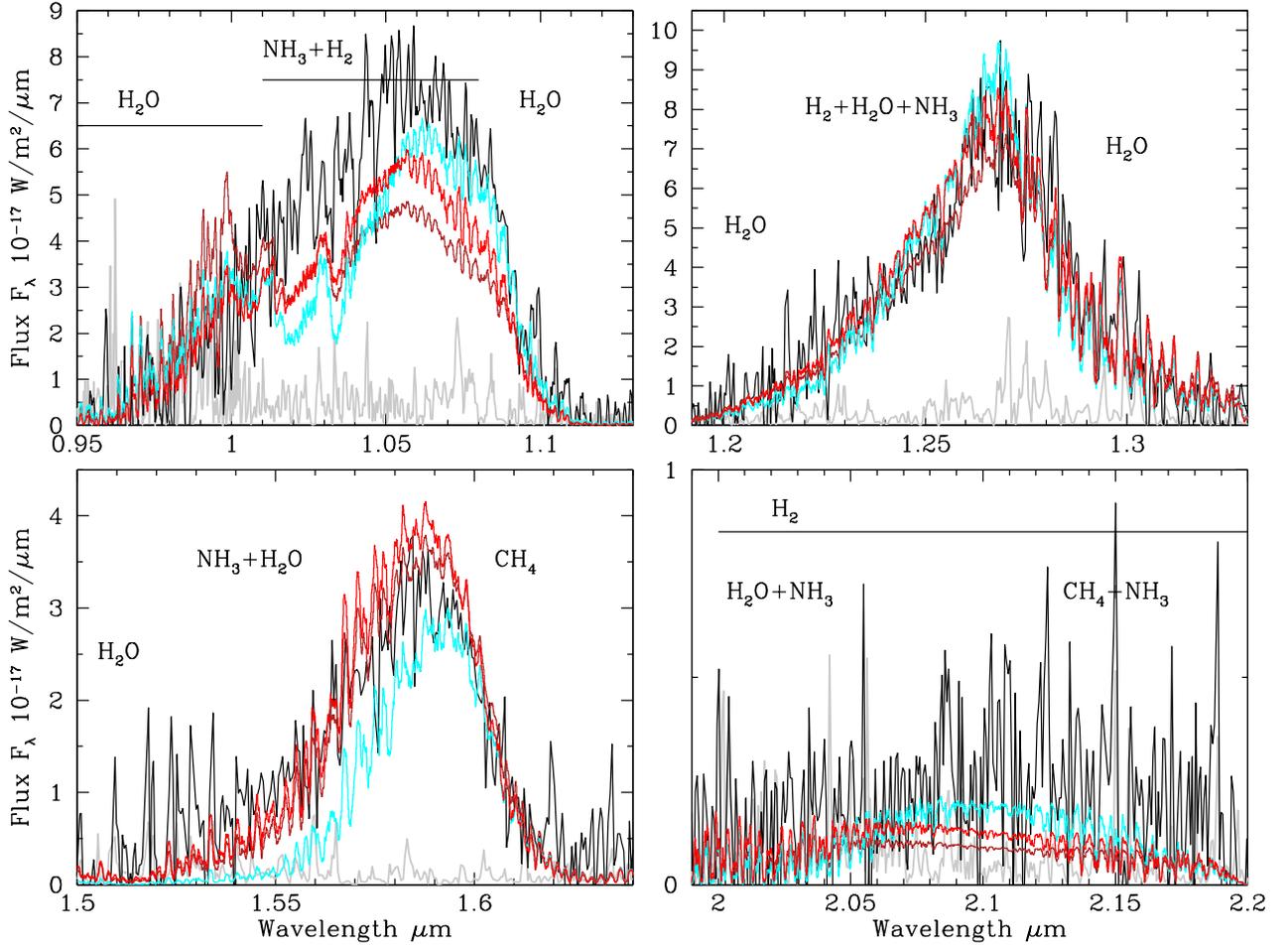}
\caption{Comparison of the observed 
WISEPC J121756.91+162640.2B  spectrum (black, Leggett et al. 2014, with 3-pixel boxcar smoothing) to T15 solar metallicity synthetic spectra smoothed to match the plotted data resolution:  $T_{\rm eff} = 450$K log $g = 4.8$  log $K_{\rm zz} = 6$ (dark red); $T_{\rm eff} = 450$K log $g = 4.5$ log $K_{\rm zz} = 6$ (red); and equilibrium chemistry $T_{\rm eff} = 450$K log $g = 4.5$  (cyan). The gray line is the uncertainty in the observational data. The model spectra have been scaled to the distance and radius of the target. The radius is supplied by evolutionary models for the given temperature and gravity (Saumon \& Marley 2008). The trigonometric parallax distance is  $10.1^{+1.9}_{-1.4}$ pc (Dupuy \& Kraus 2013);
for the  equilibrium and non-equilibrium $T_{\rm eff} = 450$K log $g = 4.5$ (red and cyan) models the distance has been adjusted slightly for a better match, to 11.3 pc. Principal absorbers are indicated.
\label{fig7}}
\end{figure}

\begin{figure}
    \includegraphics[angle=-90,width=.99\textwidth]{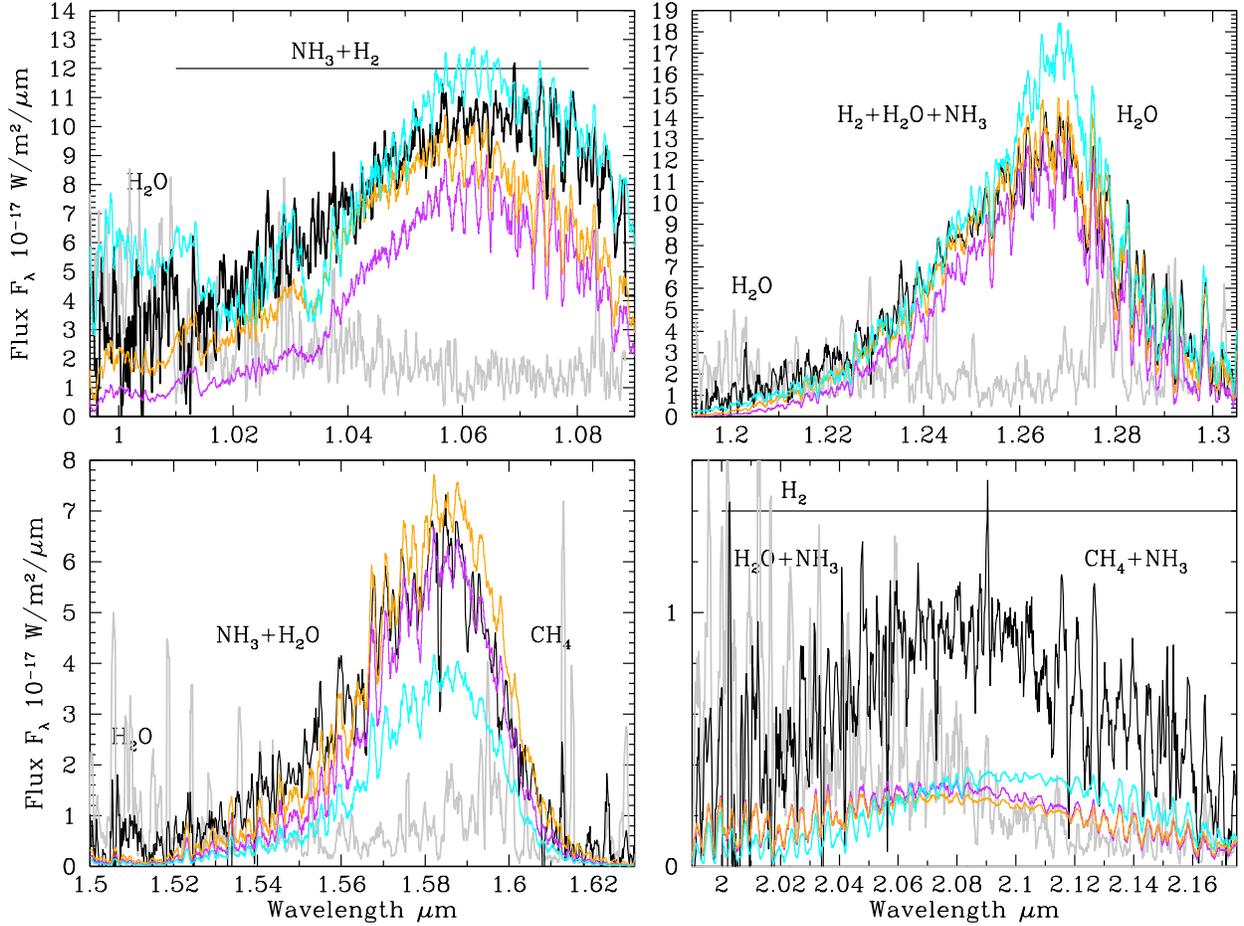}
\caption{Comparison of the observed  WISEP J173835.52$+$273258.9 spectrum (black, with 5-pixel boxcar smoothing) to T15 synthetic spectra smoothed to match the plotted data resolution: 
$T_{\rm eff} = 400$K log $g = 4.0$ log $K_{\rm zz} = 6$ [m/H] $= +0.2$ (violet);
$T_{\rm eff} = 425$K log $g = 4.0$ log $K_{\rm zz} = 6$ [m/H] $= +0.0$ (orange);
and equilibrium chemistry  $T_{\rm eff} = 450$K log $g = 4.5$ [m/H] $= +0.0$ (cyan).   The gray line is the uncertainty in the observational data. The model spectra have been scaled to the distance and radius of the target. The radius is supplied by evolutionary models for the given temperature and gravity (Saumon \& Marley 2008). The trigonometric parallax distance is 7.8 $\pm 0.6$ pc (Beichmann et al. 2014); to improve the match the 
$T_{\rm eff} = 400$K log $g = 4.0$ log $K_{\rm zz} = 6$ [m/H] $= +0.2$ (violet) model has been scaled to the low end of this range, 7.2 pc, and the  equilibrium chemistry  $T_{\rm eff} = 450$K, log $g = 4.5$ [m/H] $= +0.0$ model (cyan) has been scaled towards the high end of this range, 8.2 pc. Principal absorbers are indicated. 
\label{fig8}}
\end{figure}

\begin{figure}
    \includegraphics[angle=-90,width=.99\textwidth]{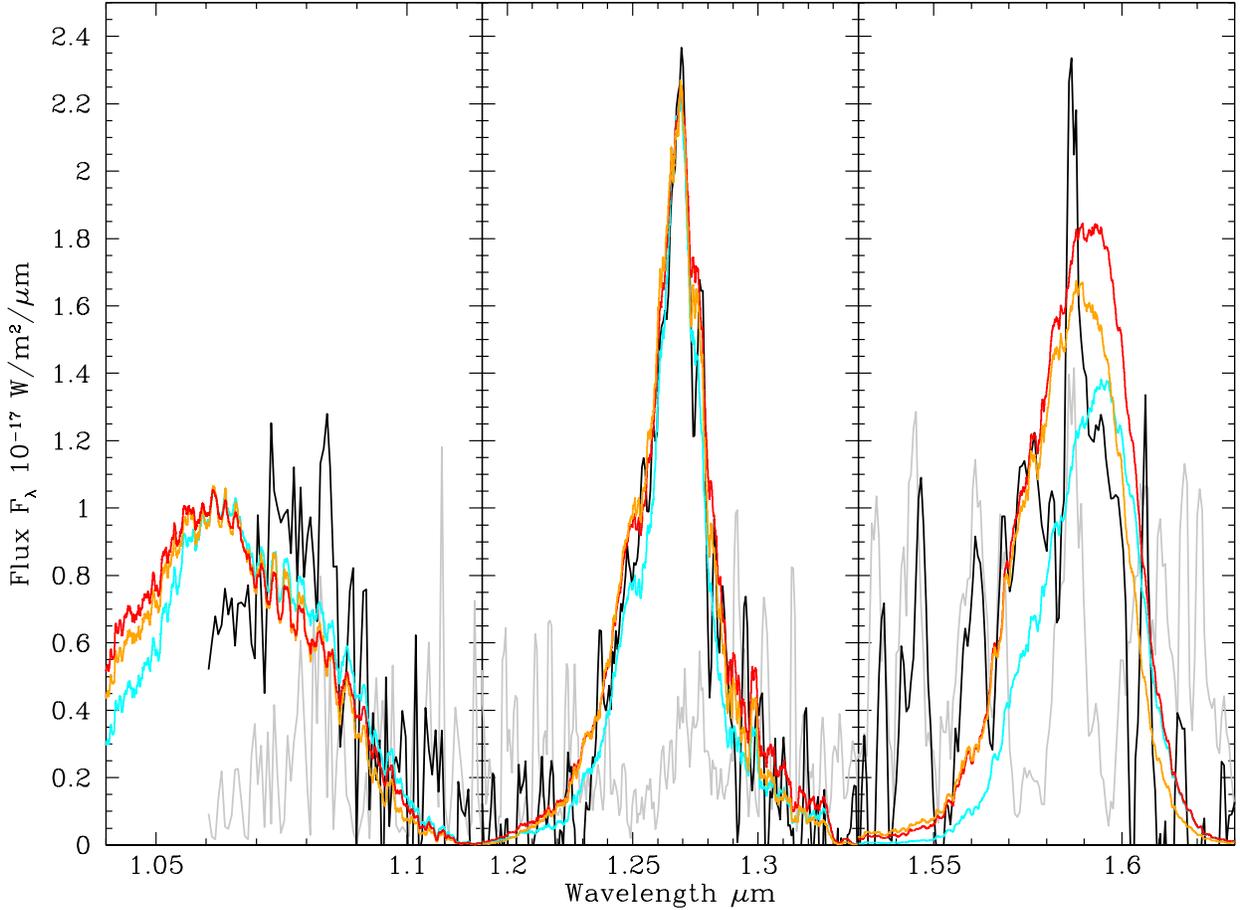}
\caption{Comparison of the observed  WISE J035000.32$-$565830.2 spectrum (black, smoothed with a 3 pixel boxcar in the central panel, and 5 pixel in the right panel) 
 to T15 solar metallicity synthetic spectra smoothed to match the plotted data resolution: 
$T_{\rm eff} = 350$K log $g = 4.5$ log $K_{\rm zz} = 6$ (red); 
$T_{\rm eff} = 350$K log $g = 4.0$ log $K_{\rm zz} = 6$ (orange); 
and equilibrium chemistry  $T_{\rm eff} = 350$K log $g = 4.5$ (cyan).
 The gray line is the uncertainty in the observational data.
The model spectra have been scaled to the distance and radius of the target. 
The radius is supplied by evolutionary models for the given temperature and gravity (Saumon \& Marley 2008). For this object the preliminary parallax distance of 2.9 -- 4.1 pc
by Marsh et al. (2013) appears more uncertain than estimated and we have allowed the distance to be greater than 4.1 pc. The distances implied by the fits shown in the plot range from 5.3 pc for the $T_{\rm eff} = 350$K log $g = 4.5$ equilibrium 
and non-equilibrium models to 6.3 pc for the  $T_{\rm eff} = 350$K log $g = 4.0$ model.
See Figure 7 for absorber identification.
\label{fig9}}
\end{figure}

\clearpage

\begin{deluxetable}{lcccc}
\tablewidth{0pt}
\tablecaption{Observational Data and Estimated Properties}
\tablehead{ 
\colhead{Property}  & \multicolumn{2}{c}{WISE J035000.32$-$565830.2} &  \multicolumn{2}{c}{WISEP J173835.52$+$273258.9} \\
\colhead{} & \colhead{Value}  & \colhead{Reference}  & \colhead{Value}  & \colhead{Reference} \\
}
\startdata
Spectral Type & Y1 & K12 & Y0 & C11\\
$M - m$(err) & 2.32(0.37)\tablenotemark{a} & M13 &  0.54(0.17)  & B14 \\
$Y_{\rm MKO}$(err) & 21.62(0.12) & L15 & 19.79(0.07) & this work \\
$J_{\rm MKO}$(err) &  22.09(0.12) & L15 &  19.63(0.05)  & this work \\
$H_{\rm MKO}$(err) & 22.51(0.20) & L15 & 20.24(0.08) & this work \\
$K_{\rm MKO}$(err) & \nodata &\nodata  & 20.58(0.10) & L13 \\
Ch.1(3.6~$\mu$m)$_{\rm IRAC}$ & 17.84(0.03) & L15 & 16.87(0.03) & L13 \\
Ch.2(4.5~$\mu$m)$_{\rm IRAC}$ & 14.61(0.03) & L15 & 14.42(0.03) & L13 \\
W1(3.4~$\mu$m)$_{WISE}$ &  \nodata & \nodata & 17.71(0.16) &     AllWISE \\
W2(4.6~$\mu$m)$_{WISE}$ & 14.75(0.04) &  AllWISE &14.50(0.04)  &   AllWISE \\
W3(12~$\mu$m)$_{WISE}$ & 12.33(0.28) & AllWISE & 12.45(0.40) &   AllWISE \\
$T_{\rm eff}$ K & 350 $\pm$ 25 & this work & 425 $\pm$ 25 & this work \\ 
log $g$ cm s$^{-2}$& 4.00 $\pm$ 0.25 & this work & 4.00 $\pm$ 0.25 & this work \\ 
Mass Jupiter & 3 -- 9  & this work & 3 -- 9 & this work \\ 
Age Gyr & 0.3 -- 3  & this work & 0.15 -- 1 & this work \\ 
\enddata
\tablenotetext{a}{This preliminary parallax results in an unrealistically faint absolute magnitude; the spectral type and model fits shown here suggest that $M - m = 1.00 \pm 0.40$.}
\tablecomments{References are: Beichman et al. 2014; Cushing et al. 2011; Kirkpatrick et al. 2012, 2013; Leggett et al. 2013, 2015;  Marsh et al. 2013.}
\end{deluxetable}

\begin{deluxetable}{lll}
\tabletypesize{\footnotesize}
\tablewidth{0pt}
\tablecaption{Brown Dwarf Identifications for Figure 4}
\tablehead{ 
\colhead{Short Name}  & \colhead{Full Name}  &    \colhead{Discovery Reference} \\
}
\startdata
W0146AB & WISE J014656.66$+$423410.0 & Kirkpatrick et al. 2012; Dupuy, Liu \& Leggett 2015 \\
W0350 & WISE J035000.32$-$565830.2  & Kirkpatrick et al. 2012 \\
W0359 & WISE J035934.06$-$540154.6 & Kirkpatrick et al. 2012 \\
W0410 & WISEP J041022.71$+$150248.5 & Cushing et al. 2011 \\
W0535 & WISE J053516.80$-$750024.9 & Kirkpatrick et al. 2012 \\
W0647 & WISE J064723.23$-$623235.5 & Kirkpatrick et al. 2013 \\
W0713 & WISE J071322.55$-$291751.9 & Kirkpatrick et al. 2012 \\
W0734 & WISE J073444.02$-$715744.0 & Kirkpatrick et al. 2012 \\
WD0806B & WD 0806$-$661B & Luhman, Burgasser \&  Bochanski, 2011 \\
W0811 & WISE J081117.81$-$805141.3 & Kirkpatrick et al. 2012 \\
W0855 & WISE J085510.83$-$071442.5 & Luhman 2014 \\
W1217AB &   WISEPC J121756.91+162640.2  & Kirkpatrick et al. 2012, Liu et al. 2012 \\
W1405 & WISEPC J140518.40$+$553421.5 & Cushing et al. 2011 \\
CFBD1458AB &   CFBDSIR J145829+101343 & Delorme et al. 2010, Liu et al. 2011 \\
W1541 & WISEP J154151.65$-$225025.2 & Cushing et al. 2011 \\
W1738 & WISEP J173835.52$+$273258.9 & Cushing et al. 2011 \\
W1828 & WISEP J182831.08$+$265037.8 & Cushing et al. 2011 \\
W2056 & WISEPC J205628.90$+$145953.3 & Cushing et al. 2011 \\
W2220 & WISE J222055.31$-$362817.4 & Kirkpatrick et al. 2012 \\
\enddata
\end{deluxetable}

\end{document}